\documentclass[referee]{aa} 
\usepackage{graphicx}
\usepackage{txfonts}
\usepackage[]{natbib}
\begin{document}
\title{Observations of H1426+428 with  HEGRA}
\subtitle{Observations in 2002  and reanalysis of 1999\&2000 data}
\author{
F.~Aharonian\inst{1},
A.~Akhperjanian\inst{7},
M.~Beilicke\inst{4},
K.~Bernl\"ohr\inst{1},
H.-G.~B\"orst\inst{5},
H.~Bojahr\inst{6},
O.~Bolz\inst{1},
T.~Coarasa\inst{2},
J.L.~Contreras\inst{3},
J.~Cortina\inst{10},
L.~Costamante\inst{1},
S.~Denninghoff\inst{2},
M.V.~Fonseca\inst{3},
M.~Girma\inst{1}
N.~G\"otting\inst{4},
G.~Heinzelmann\inst{4},
G.~Hermann\inst{1},
A.~Heusler\inst{1},
W.~Hofmann\inst{1},
D.~Horns\inst{1},
I.~Jung\inst{1},
R.~Kankanyan\inst{1},
M.~Kestel\inst{2},
A.~Kohnle\inst{1},
A.~Konopelko\inst{1},
H.~Kornmeyer\inst{2},
D.~Kranich\inst{2},
H.~Lampeitl\inst{4},
M.~Lopez\inst{3},
E.~Lorenz\inst{2},
F.~Lucarelli\inst{3},
O.~Mang\inst{5},
D.~Mazine\inst{4},
H.~Meyer\inst{6},
R.~Mirzoyan\inst{2},
A.~Moralejo\inst{3},
E.~Ona-Wilhelmi\inst{3},
M.~Panter\inst{1},
A.~Plyasheshnikov\inst{1,8},
J.~Prahl\inst{4},
G.~P\"uhlhofer\inst{1},
R.~de\,los\,Reyes\inst{3},
W.~Rhode\inst{6},
J.~Ripken\inst{4},
G.~Rowell\inst{1},
V.~Sahakian\inst{7},
M.~Samorski\inst{5},
M.~Schilling\inst{5},
M.~Siems\inst{5},
D.~Sobzynska\inst{2,9},
W.~Stamm\inst{5},
M.~Tluczykont\inst{4},
V.~Vitale\inst{2},
H.J.~V\"olk\inst{1},
C.~A.~Wiedner\inst{1},
W.~Wittek\inst{2}}
\institute{
Max-Planck-Institut f\"ur Kernphysik, Postfach 103980, D-69029 Heidelberg, Germany   
\and Max-Planck-Institut f\"ur Physik, F\"ohringer Ring 6, D-80805 M\"unchen, Germany
\and Universidad Complutense, Facultad de Ciencias F\'{\i}sicas, Ciudad Universitaria, E-28040 Madrid, Spain
\and Universit\"at Hamburg, Institut f\"ur Experimentalphysik, Luruper Chaussee 149, D-22761 Hamburg, Germany
\and Universit\"at Kiel, Institut f\"ur Experimentelle und Angewandte Physik, Leibnizstra{\ss}e 15-19, D-24118 Kiel, Germany
\and Universit\"at Wuppertal, Fachbereich Physik, Gau{\ss}str.20, D-42097 Wuppertal, Germany
\and Yerevan Physics Institute, Alikhanian Br. 2, 375036 Yerevan, Armenia
\and On leave from Altai State University, Dimitrov Street 66, 656099 Barnaul, Russia
\and Home institute: University Lodz, Poland
\and Now at Institut de F\'{\i}sica d'Altes Energies, UAB, Edifici Cn, E-08193, Bellaterra (Barcelona), Spain
}

\authorrunning{Aharonian et al.}

\offprints{D. Horns} \mail{Dieter.Horns@mpi-hd.mpg.de} \date{Received ;
accepted} \def\mscw{\textit{mscw}}
\abstract{ The HEGRA system of imaging air Cherenkov telescopes has been
used to observe the BL Lac object H1426+428 ($z=0.129$) for 217.5~hours
in 2002. In this data set alone, the source is detected at a confidence
level of $5.3~\sigma$, confirming this object as a TeV source. The
overall flux level during the observations in 2002 is found to be a
factor of $\approx 2.5$ lower than during the previous observations by
HEGRA in 1999\&2000. A new spectral analysis has been carried out,
improving the signal-to-noise ratio at the expense of a slightly increased
systematic uncertainty and reducing the relative energy
resolution to $\Delta E/E\le 12\,\%$ over a wide range of energies.  The
new method has also been applied to the previously published data set
taken in 1999 and 2000, confirming the earlier claim of a flattening of
the energy spectrum between 1 and 5~TeV. The data set taken in 2002
shows again a signal at energies above 1~TeV. We combine the energy
spectra as determined by the CAT and VERITAS groups with our reanalyzed
result of the 1999\&2000 data set and apply a correction to account for
effects of absorption of high energy photons on extragalactic background
light in the optical to mid infrared band.  The shape of the inferred
source spectrum is mostly sensitive to the characteristics of the
extragalactic background light between wavelengths of  1 and 15~$\mu$m.
\keywords{BL Lacertae objects: individual H1426+428 -- Gamma-rays:
observations -- diffuse radiation -- intergalactic medium} }
 \maketitle

\section{Introduction}

 The extragalactic objects H1426+428 \citep{1989ApJ...345..140R} and Mkn~501
are the only established members of the class of extreme synchrotron BL Lac
objects \citep{2001A&A...371..512C} detected at TeV energies. Generally, these
objects are characterized by a flat X-ray spectrum with $f_{\nu}\propto
\nu^{-\alpha}$ with $\alpha<1$ extending into the hard X-ray energy band. In the
framework of synchrotron self-Compton (SSC) models, these objects are of
particular interest for observations at TeV energies, because the inverse
Compton peak of the broad band spectral energy distribution (SED) is expected
to be located at these high energies. The measurement of the position of the
inverse Compton peak is hampered by the effect of pair-production of high
energy photons with low energy (optical and infrared) photons of the
extragalactic background light \citep{nikishov,gs1967,sds}.  For the given
red-shift $z=0.129$ of H1426+428 the optical depth exceeds unity  even for
energies of a few 100~GeV. Any detection of a signal at TeV energies translates
directly into a high luminosity of the source. 

 The detection and spectral measurements of H1426+428 by the CAT, and
VERITAS groups \citep{2002A&A...391L..25D,2002ApJ...580..104P} indicate
a steep spectrum between 250~GeV and 1~TeV (well described by a power
law $dN/dE\propto E^{-n}$ with a photon index of $n=3.66\pm0.41$ for the
CAT and $n=3.54\pm0.27$ for the VERITAS measurements)  whereas at higher
energies, a detection has been claimed only  by the HEGRA group based
upon observations carried out in 1999 and 2000 with $n=2.6\pm0.6$
\citep[][in the following Paper
I]{2002A&A...384L..23A}\defcitealias{2002A&A...384L..23A}{Paper~I}.  The
spectrum shows a flattening at energies above 1~TeV that is consistent
with the expected signature of absorption as discussed in Paper~I. A
deep follow-up observation (217.5~hrs) has been carried out in 2002 to
investigate the energy spectrum with improved statistics. Results of a
reanalysis of the original data set and the results from the recent
observations of 2002 are reported.  A possible excess from H1426+428
with a significance of 4.1 $\sigma$ has been seen with the
AIROBICC-Scintillator arrays of HEGRA in the data from 1994/95
corresponding to an integral flux of $\rm  (5.7 \pm 1.3)\cdot 10^{-13}
cm^{-2} s^{-1}$ above $ \rm E_{thresh} = 21$~TeV  \citep{prahl}.  The
interpretation and the relation of this excess to data presented here is
not clear at present.

\section{Analysis technique}

The spectral analysis applied to data from the Crab-Nebula
\citep{2000ApJ...539..317A}, Mkn~421 \citep{2002A&A...393...89A}, and
Mkn~501 \citep{1999A&A...349...11A}  benefits from a signal-to-noise
ratio ($S/N$) which is usually $S/N \gg 1$. Therefore, relaxed cuts
(compared to signal search optimized cuts) on the \textit{mean scaled
width (\mscw)} parameter and the angular separation from the source
($\theta$) assures good photon statistics with little systematic
effects. However, for weaker sources, the signal-to-noise ratio $S/N$
needs to be improved.  The relaxed spectral cuts result in a poor
$S/N=0.1$ for H1426+428.  A tighter cut on \mscw\, and a smaller angular
search bin increases the $S/N$ to $0.3$ reducing at the same time the
photon efficiency to the level of $50~\%$. See Table~\ref{table:cuts}
for a summary of the cuts applied to select the data set, events, and
images used in this analysis.

 The algorithm for reconstructing the energy of the primary photon has
been improved following the methods suggested in
\citet{2000APh....12..207H}. The relative energy resolution for
individual events has been improved to $\Delta E/E\le 12\,\%$ for a wide
range of energies. For energies above 1~TeV, the relative energy
resolution is as low as $9\,\%$, slightly increasing for lower energies.
For the threshold region above 500~GeV, the bias of the energy
reconstruction method has been reduced substantially with respect to the
conventional methods. The main contribution for this improvement stems
from a correction for the position of the shower maximum. Low energy air
showers triggering the telescopes tend to penetrate deeper into the
atmosphere and therefore are reconstructed at higher energies. This
strong bias is overcome and allows for a more reliable energy
reconstruction at threshold energies. The reconstruction of the core
position has been improved by assuming a point-like source position
\citep{1999APh....12..135H}.  The improved energy resolution reduces
spill-over effects distorting the reconstructed energy spectrum in the
presence of sharp cut-off features in the initial spectrum.

 The scaling of individual image widths to the expectation for simulated
showers has been modified in order to benefit from the improved core
reconstruction. This results in a slightly narrower distribution of the
\mscw -parameter  with respect to the conventional core reconstruction
method. We have cross-checked the background subtracted distribution of
\mscw~ for strong sources with Monte Carlo predictions and see a very
good agreement of the distribution between data and simulation.

\begin{table}
 \caption{%
Listed are data, image- and event-selection criteria. $n_{\mathrm{tel}}$
and $n_{\mathrm{image}}$ indicating the number of active telescopes
participating in the data taking and the number of images used for the
event reconstruction respectively. $R_{\mathrm{meas.}}$ and
$R_{\mathrm{CR}}$ are the average measured and expected trigger rate of
the telescopes.  $r_{\mathrm{core}}$ is the distance of the shower
impact point to the position of the central telescope. The altitude
angle above horizon is indicated as $alt$, whereas $\theta^2$ is the
squared angular distance of the shower direction to the source
direction. The $\gamma$-hadron separating quantity $mscw$ is further
explained in the text. The image parameters $distance$ and $size$
describe the position of the shower image centroid with respect to the
center of the camera and the total amount of light (in photo electrons,
p.e.) measured in the image.  
}

 \label{table:cuts}
 \begin{tabular}{lll}
 \hline\hline
 Data selection & Event selection & Image selection \\
 \hline 
 $n_{\mathrm{tel}}\ge 3$                                 &  $n_{\mathrm{image}}\ge2$  & $distance<1.7^\circ$ \\
 $|1-R_{\mathrm{meas.}}/\langle R_{\mathrm{CR}}(\theta,n_{\mathrm{tel}})\rangle|$ & $r_{\mathrm{core}}<200$~m& $size>40$~p.e.        \\
$<0.2$                                                         & $alt>45^\circ$  & \\
                                                         & $\theta^2<0.014~ \mathrm{deg}^2$     &  \\
                                                         & $mscw<1.1$            &  \\
\hline
 \end{tabular}
\end{table}

 The complete analysis chain has been tested with data taken on the
Crab-Nebula.  The results on flux and spectral shape as published in
\citet{2000ApJ...539..317A} are reproduced. The systematic error on the
reconstructed photon index is estimated to increase from $0.05$ in the
case of relaxed cuts analysis to $0.08$ for the analysis presented here.
However, given the dominating statistical error on low-flux sources,
this can be tolerated.

\begin{table} 
\caption{Listed are for the two data sets: the individual
\label{table:summary} event statistics ($N_{\mathrm{on}}$,
$N_{\mathrm{off}}$, and
$N_s=N_{\mathrm{on}}-\Omega_{\mathrm{on}}/\Omega_{\mathrm{off}}\cdot
N_{\mathrm{off}}$), the rate of excess events per hour ($\dot
N_\mathrm{s}$), and the significance $S$ calculated using Eq.~17 of
\citet{1983ApJ...272..317L}. The quantity
$\Omega_{\mathrm{on}}/\Omega_{\mathrm{off}}$ is the normalization for
the background estimate derived by taking the ratio of the solid angle
subtended by the two separate regions.}
\centerline{
\begin{tabular}{lcc}
\hline\hline
                             & reanalyzed &    \\
                             & 1999-2000 & 2002\\
\hline
$T_{\mathrm{obs}}\,[{\rm hrs}]$       & 42.6          & 217.5\\
$\langle alt \rangle\,[^\circ]$      & 70.8          & 68.4 \\
median($alt$)        $[^\circ]$     & 73          & 70.4 \\
$N_{\mathrm{on}} $                    & 309           & 1095  \\
$N_{\mathrm{off}}$                    & 1447          & 6425   \\
$\Omega_{\mathrm{on}}/\Omega_{\mathrm{off}}$                     & 1/7           & 1/7    \\
$N_{\mathrm{s}}  $                    & 102.3$\pm16.6$  & 177.1$\pm33.5$  \\
$\dot{N_{\mathrm{s}}}[1/\mathrm{hr}]$ &  2.4$\pm0.4$  & $0.81\pm0.15$  \\
$S\,[\sigma]$                &  6.1         & 5.3  \\
\hline
\end{tabular}
}
\end{table}

\begin{table*}
\caption{Event statistics for individual energy bins. \label{table:spec}}
\centerline{
\begin{tabular}{lcccc|cccc}
\hline \hline
                & \multicolumn{4}{c}{1999-2000} & \multicolumn{4}{c}{2002} \\
   E            & $N_\mathrm{on}$ & $N_\mathrm{off}$ & S & Flux                            & $N_\mathrm{on}$ & $N_\mathrm{off}$ & S & Flux  \\
$[\mathrm{TeV}]$&          &           & $[\sigma]$  
& $[10^{-13}$ph~cm$^{-2}$~s$^{-1}$~TeV$^{-1}]$&  & & $[\sigma ]$&
  $[10^{-13}$ph~cm$^{-2}$~s$^{-1}$~TeV$^{-1}]$\\
\hline
0.78 & 122 & 576&  3.78 &$187.3 \pm  78.7 $ &  366 & 2187 & 2.75 & $12.3\pm37.8$\\
1.48 & 86 & 456 &  2.29 &$8.11 \pm   3.51 $ &  394 & 2449 & 2.16 & $2.02\pm1.40$  \\
2.82 & 42 & 179 &  2.75 &$1.92  \pm   0.75 $ &  159 & 826 & 3.33 & $0.66\pm0.24$ \\
5.37 & 16 & 46  &  2.83 &$0.47 \pm   0.20$ &  60  & 226  & 4.00 & $0.18\pm0.05$\\
10.2 &  4 & 12  &  1.36 &$0.036 \pm   0.034$ &   7  & 58   & -0.43 & $<0.03$ (99\,\% c.l.)       \\
\hline
\end{tabular}
}
\end{table*}

\begin{figure}
\centerline{\includegraphics[width=0.9\linewidth]{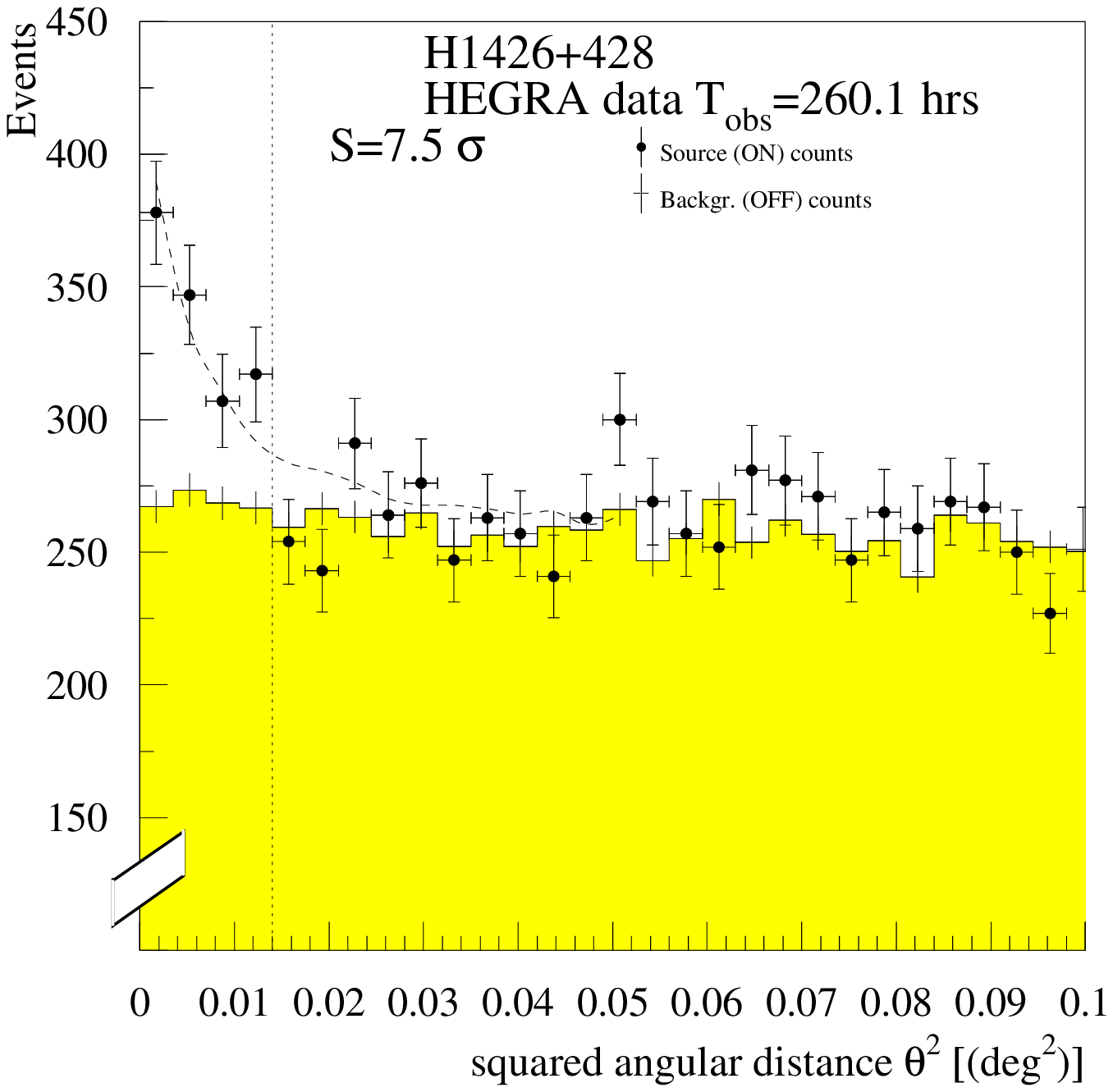}}
\caption{\label{Fig1} 
 The number of events reconstructed as a function of the angular separation to 
the source position:
After applying a cut on $mscw<1.1$ and $\theta^2<0.014$~deg$^2$ (dashed line) to reject
background events, an excess with a significance of $S=7.5\,\sigma$
is observed from the direction of H1426+428 for the combined data set encompassing 260.1~hrs of
observations. 
 The dashed histogram
\label{fig:theta}
indicates the expected shape of a signal from a point-like source determined from
a scaled-down signal of the Crab-Nebula. 
}
\end{figure}

\section{Results}

 The data taken on H1426+428 in 1999 and 2000 with the HEGRA system of
imaging air Cherenkov telescopes have been reanalyzed applying the
modified and improved methods described above. The data set has been
described in our earlier publication (see Table 1 in Paper~I for details
on the dates of observations).  The signal has been recovered with a
slightly higher significance than previously claimed increasing the
significance from $S(\mathrm{Paper\, I})=5.8~\sigma$ to $S=6.1~\sigma$.
Note, we have not performed a sensitivity-optimized source search. The
cuts are chosen to perform a spectral analysis (which includes a
constraint on the core distance).  The deep exposure of 217.5~hrs in
2002 confirms the object as a $\gamma$-ray emitter with a significance
for the signal of $S=5.3~\sigma$ (see Table~\ref{table:summary}). The
combined data set with an observation time of $260.1$~hrs results in
$S=7.5~\sigma$ (see Fig.~\ref{fig:theta} for the distribution of the
arrival directions).

\begin{figure}
\includegraphics[width=0.8\linewidth]{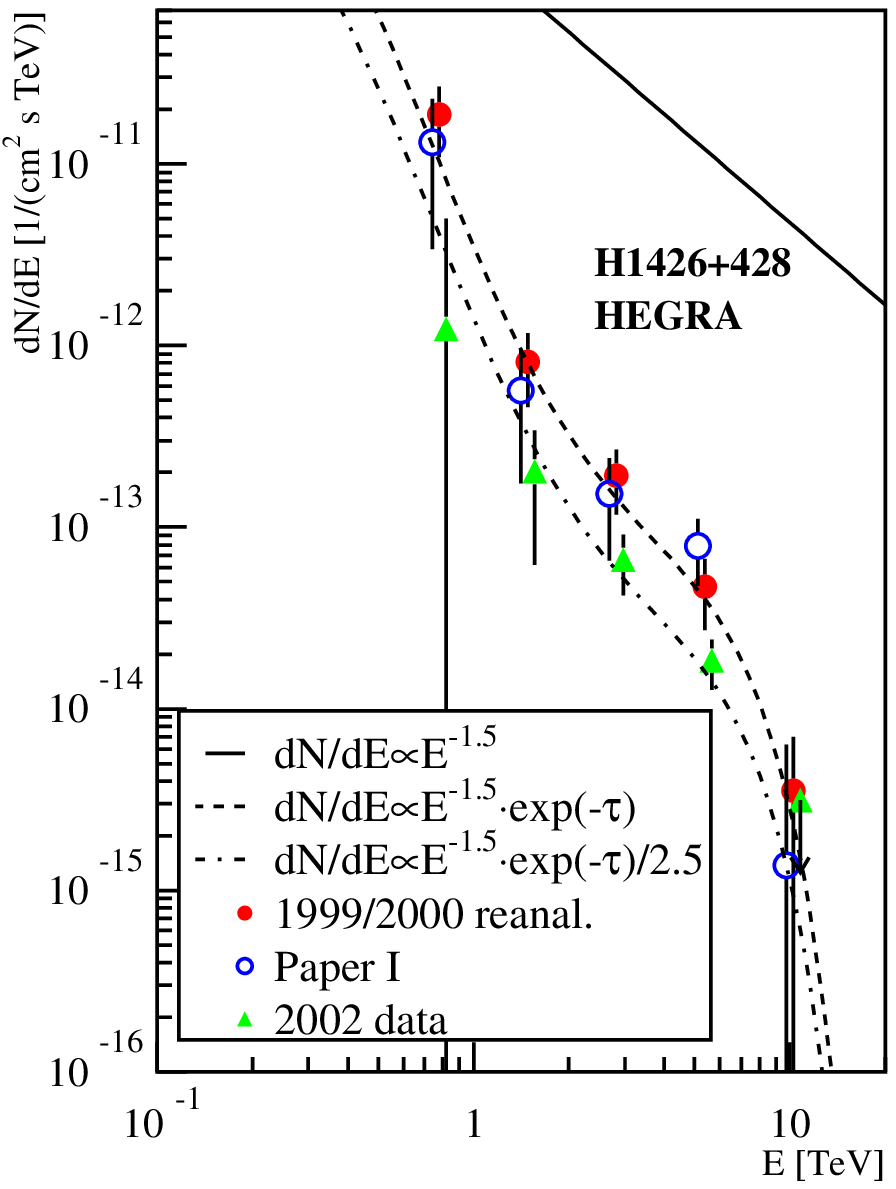}
\caption{\label{Fig:spectrum}
The differential energy spectra for the different data sets:
 The reanalyzed data set taken in the years 1999 and 2000
reproduces the initially published spectrum with smaller error bars. The solid and 
dashed curves indicating a possible intrinsic source spectrum and the effect of 
absorption are calculated as those shown later in  Fig.~\ref{fig:spec_comp}a. 
The data set taken in 2002 is
compatible with the 1999/2000 data  with a  flux level reduced by a factor of 2.5 (dot-dashed curve).
 See also Table~\ref{table:spec} for
the event statistics and differential flux values for the individual energy bins. The displayed 
points are slightly shifted in energy with respect to each other to allow for easier reading.}
\end{figure}

 The reconstructed energy spectrum of the reanalyzed data set from
1999/2000 agrees well with the previously published result (see
Fig.~\ref{Fig:spectrum}).  Given the quite different approach in event
selection and reconstruction technique applied in the present analysis,
this result demonstrates the negligible influence of systematic effects
caused by different analysis techniques.  

The recent deep exposure obtained in 2002  shows a decrease in the
time-averaged flux by a factor of $2.5$ with comparison to the
observations in 1999 and 2000 (see Fig.~\ref{Fig:spectrum}). The
decrease in flux observed at high energies is similar to the decrease of
the time-averaged X-ray flux as measured by the all-sky-monitor (ASM)
on-board the Rossi X-ray timing explorer (RXTE).  The average of the
daily rates of the ASM provided by the ASM/RXTE team for the nights of
HEGRA observations in 1999 and 2000 is $0.33\pm0.08$~counts/sec.  For
the nights of HEGRA observations in 2002,  the X-ray rate as measured by
the ASM drops by a factor of 1.7 to  $0.19\pm0.04$~counts/sec.

\begin{figure}
\includegraphics[width=\linewidth]{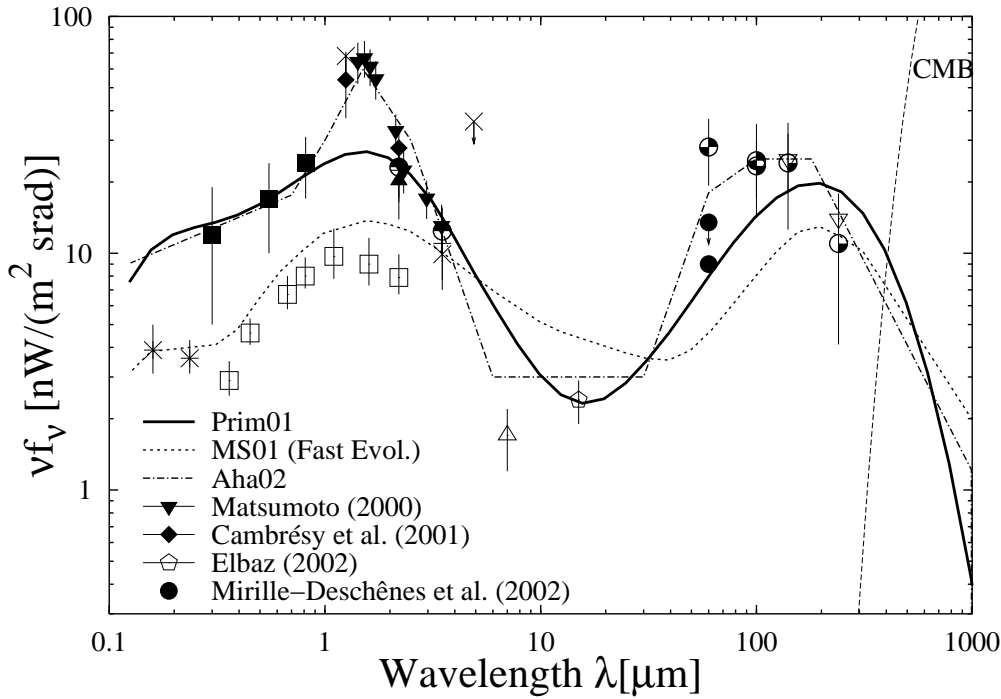}
\caption{The spectral energy distribution (SED) of the 
of the extragalactic background light (EBL). The compilation of measurements
and limits  between ultraviolet and
far infrared are taken from \citet{2001ARA&A..39..249H}. 
 Recently published results have been added:
the satellite based measurements between 1 and 4~$\mu$m \citep{matsumoto},
 the combined DIRBE/2MASS analyses claiming a detection at $1.25$ and
2.2~$\mu$m \citep{2001ApJ...555..563C},  a new analyses of the 15~$\mu$m 
data taken with the ISO satellite \citep{2002A&A...384..848E},  a model-dependent measurement, and
a firm upper-limit at $60~\mu$m based upon fluctuation measurements in IRAS data \citep{2002A&A...393..749M}.
Additionally, three largely different models for 
the spectral energy distribution (SED) of the EBL are included based upon 
\citet{prim01} (Prim01, solid curve), \citet{2001ApJ...555..641M} (MS01 ``fast evolution'', dashed curve), and
a phenomenological description of the data (dash-dotted curve, labelled Aha02), 
interpolating between the measured points (Paper I).
\label{fig:IR-SED}}
\end{figure}

\begin{figure*}
\includegraphics[width=\linewidth]{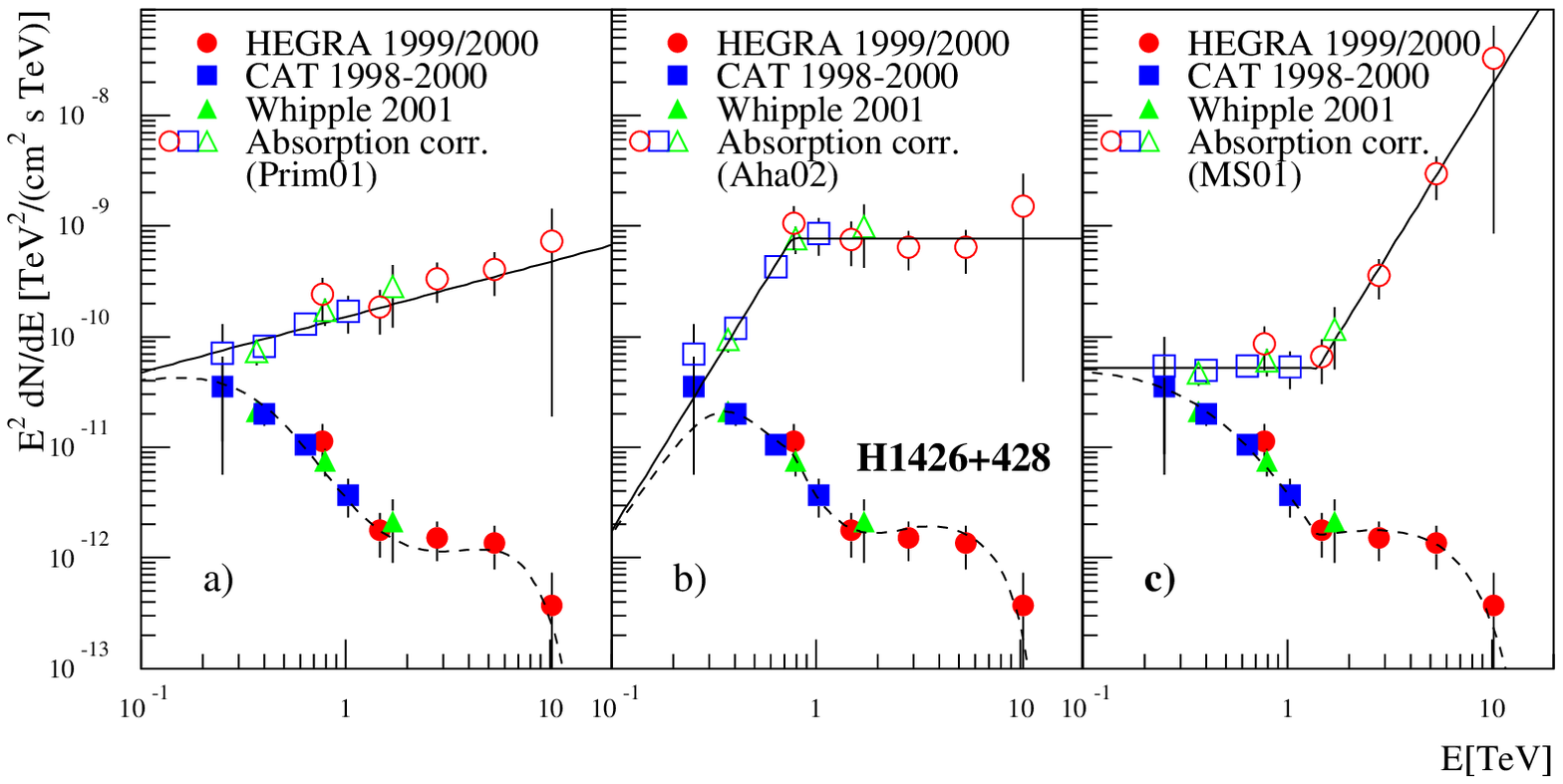}
\caption{
	The three panels (a-c) display the observed differential flux values multiplied by 
$E^2$. The filled symbols indicate the combined observed spectra,  whereas 
		the open symbols indicate the intrinsic spectrum  inferred from the observed data
		by applying a model-dependent absorption correction. The solid lines are
		fits of a power law (a) and broken power law (b-c) functions to the intrinsic
		spectrum. The lower dashed curves are the absorption corrected fit functions. 
		\textbf{a)} 
	Based upon the Prim01 EBL (see also Fig.~\ref{fig:IR-SED}), the absorption corrected
		spectrum follows a power law $E^2\,dN/dE \propto$E$^{0.5}$.
		\textbf{b)} The fast increase of absorption predicted above 200~GeV for the Aha02 EBL leads to 
		a steeply rising ($E^2 dN/dE\propto$E$^{3}$) source spectrum up to $\approx700$~GeV. Data at higher energies
		are consistent (as claimed in Paper~I) with 
		a $E^2\,dN/dE\propto const.$.
		\textbf{c)} Applying an absorption correction according to the MS01 model: 
		We note a marked upturn of the corrected spectrum at energies above $\approx 1$~TeV
		with $E^2\, dN/dE\propto E^3$. Data below energies of 1~TeV follow
		$E^2\,dN/dE \propto const.$.
		\label{fig:spec_comp}
}
\end{figure*}

\section{Discussion}

In order to correct for the effect of absorption of the energetic
photons by interactions with the extragalactic background light (EBL) in
the near (1-3.5~$\mu$m, NIR) to mid (4.5-20~$\mu$m, MIR) infrared, three
largely different models for the spectral energy distribution (SED) of
the EBL have been used to calculate the opacity $\tau(E,z)$. 

The calculations of the optical depth $\tau$ performed here apply a
cosmological model  with  a flat cold dark matter ($\rho_m=0.3~\rho_c$,
$\rho_c=3H_0^2/(8\pi G)$ is the critical density) dominated universe
with a non-vanishing cosmological constant $\Lambda$
($\rho_\Lambda=0.7~\rho_c$) and
$H_0=60~\mathrm{km}~\mathrm{s}^{-1}~\mathrm{Mpc}^{-1}$. Evolutionary
effects of the EBL  have been neglected except for the expansion of the
universe.  

The three models are shown in Fig.~\ref{fig:IR-SED} together
with a selection of measurements and limits on the EBL from the ultra
violet to the far infrared wavelength bands.  The contribution of the
far infrared part of the EBL ($\lambda>100~\mu$m) and the cosmic
microwave background to the opacity is negligible, but has been included
in the calculations of $\tau(z,E)$.  The model presented by
\citet{2001ApJ...555..641M,2002ApJ...566..738D} (in the following MS01)
is based upon a backward-evolution approach with a fast evolving
luminosity function between red-shifts 0 and 1.3. For the sake of
clarity only one of the two alternative models described in MS01
(\textit{fast-evolution} model) is used in the following discussion.
Both models follow the same shape in the NIR and differ mainly in the
absolute level of the radiation (roughly 20\,\% lower flux in the
so-called \textit{base-line} model with respect to the
\textit{fast-evolution} model).

An alternative model \citep[][in the following Prim01]{prim01}
invokes a semi-analytical approach where the structure formation in a
hierarchical scenario is combined with an initial mass function
following a Kennicut-type distribution in order to calculate the SED of
individual galaxies including realistic dust re-emission models.
Following the approach used in Paper~I, a simple interpolation of the
EBL data (labelled Aha02) has been used as a third alternative.

The measurements and limits on the EBL indicated in
Fig.~\ref{fig:IR-SED} have been taken from \citet{2001ARA&A..39..249H}
and updated for recent measurements  in the NIR
\citep{matsumoto,2001ApJ...555..563C}, MIR \citep{2002A&A...384..848E},
and at 60~$\mu$m \citep{2002A&A...393..749M}, 100, 140, and 240~$\mu$m
\citep{lagache}. Whereas at $60~\mu$m the experimental situation is unclear
by a factor of $\approx 3$ between the different detections \citep{2000ApJ...544...81F,2002A&A...393..749M},
the claimed values for $\lambda=100~\mu$m are well consistent with each other.

The opacities
$\tau_\mathrm{Prim01}(z,E),\tau_\mathrm{Aha02}(z,E)$ have been
calculated by integrating over the respective SED whereas for the MS01
model  the prescription of the authors as given in
\citet{2002ApJ...566..738D} has been used. 

The wide energy range from 700~GeV up to 10~TeV
covered by the HEGRA energy spectrum of H1426+428 offers the unique
opportunity to probe the energy region where the source spectrum is
heavily absorbed by pair-production processes on the NIR/MIR part of the
EBL.

The lower energy spectra published by the CAT and VERITAS collaboration
\citep{2002A&A...391L..25D,2002ApJ...580..104P} above an energy
threshold of $\approx250$~GeV provide additional information on the
spectral shape at energies where the absorption effect is of less
importance than for energies above 700~GeV that are  accessible by the
HEGRA instrument.  The CAT group has performed a spectral analysis
applying a forward-folding technique. The residuals of the reconstructed
energy spectrum with respect to the forward-folded power law assumption
contain information on possible deviations from a pure power law. The
actual data points shown in Fig.~\ref{fig:spec_comp} are derived by
taking the residuals and converting them into differential flux
measurements. The caveats of using individual data points are clearly
larger errors on the data points and negligence of the correlation of
the energy bins which in turn depends on the assumed spectral shape.
However, this method allows to compare the CAT measurements with the
results of other spectral analysis methods, which suffer from the same
caveats.

The combined data sets for H1426+428 based upon measurements
with the CAT, Whipple, and HEGRA telescopes (reanalyzed 1999-2000 data
set) are presented in Fig.~\ref{fig:spec_comp}~a-c.  The 2002 data set
has not been considered here, because of  the lower flux level during
that observation. The agreement of the different measurements is very
good despite the fact that the observations were not carried out
simultaneously.  The X-ray flux as measured by the ASM averaged over the
individual observational windows for CAT ($0.2\pm0.4$~cts/sec), Whipple
($0.26\pm0.07$~cts/sec), and HEGRA ($0.33\pm0.08$~cts/sec) are in
reasonable agreement with each other.  The overlap of the HEGRA
measurements with the results from Whipple and CAT between 700~GeV to
1~TeV underlines the consistency of the different measurements.  The CAT
and VERITAS spectra are consistent with each other and no \textit{ad
hoc} normalization is required to combine the different data sets.

	After applying a correction to the observed spectrum by
multiplying the individual data points by
$\exp(\tau_\mathrm{Prim01}(z=0.129,E))$ we obtain a source spectrum
consistent with a hard power law with a photon index of $1.5$ (see solid
and dashed lines in Fig.~\ref{fig:spec_comp}a-c respectively for the
source spectrum and the observable spectrum).  A $\chi^2$-minimization
keeping the photon index fixed at $1.5$  results in  a
$\chi^2/d.o.f.=3/11$. This intrinsic source spectrum would imply a
position of the inverse Compton peak in the broad-band SED beyond
10~TeV.  The energy flux of the unabsorbed inverse Compton component
exceeds $10^{-10}$~erg~cm$^{-2}$~s$^{-1}$.  An initial modeling with
leptonic emission models of the broad band SED and specifically the hard
source spectrum at TeV energies is possible, but requires an additional
external radiation field \citep{luigi}.  Given the low X-ray energy flux
observed from this source of a few $10^{-11}$~erg~cm$^{-2}$~s$^{-1}$
during contemporaneous observations, the TeV observations strongly
suggest that the $\gamma$-ray flux dominates over the X-ray flux.  This
would violate the proclaimed BL Lac sequence \citep{1998MNRAS.299..433F}
where the energy flux from the synchrotron radiation  is expected to
be comparable or to dominate over the inverse Compton energy flux for TeV blazars.

	The  possibly large photon density of the  EBL between
$1$-$2~\mu$m as suggested in Paper~I to accommodate the NIR data by
\citet{matsumoto,2001ApJ...555..563C} results in a strong absorption for
$\gamma$ energies below 1~TeV. As a consequence of this, the inferred
source spectrum shown in Fig.~\ref{fig:spec_comp}b rises sharply
between 0.25~ and 0.7~TeV ($E^2 dN/dE \propto E^{3}$) with a subsequent
flattening to a constant value.  The inferred source spectrum obtained
by taking HEGRA data alone confirms the result as given in Paper~I where
the spectral shape of the TeV- and  X-ray data  was tentatively
interpreted as self synchrotron-Compton emission.  However, the inferred
rising lower energy part of the unabsorbed source spectrum is not
readily explained by such a model.

	In the framework of the MS01 model, the flat slope of the SED of
the EBL between the NIR and the MIR produces a rapidly increasing
optical depth for energies above 1~TeV, exceeding $\tau=5$  at 2~TeV.
Essentially the same behaviour is seen for the somewhat lower EBL level
given by the base-line evolution model.  After applying the correction
for absorption based upon the MS01 model, the implied source spectrum is
consistent with a power law type source spectrum with a photon-index of
2 up to 1~TeV (corresponds to $E^2\,dN/dE\propto const.$).  Trying to
fit the data points beyond 1~TeV with $E^2\,dN/dE\propto const.$ results
in a poor $\chi^2/d.o.f.=10/3$. A broken-power law with
$E^2\,dN/dE\propto E^3$ above 1.4~TeV is a good description of the
intrinsic spectrum ($\chi^2/d.o.f.=3.2/10$. A similar upturn is observed
for the base-line evolution model (omitted in Fig.~\ref{fig:spec_comp}
for the sake of clarity).  

This implied steep upturn in the corrected
source spectrum ($E^2 dN/dE\propto E^3$) is not easily explained by
commonly used models invoking either a leptonic or hadronic origin of
$\gamma$-ray emission of BL Lac objects. In the case of $\gamma$-ray
production from $\pi^0$-decay from monoenergetic $\pi^0$  a $E^2
dN/dE\approx E^2$ type spectrum could be produced. However, even in this
extreme case, a steep rise as it is observed here  is not explainable
(see e.g. \citet{2000A&A...354..395P}). 
  
	A sharper pile-up could be expected by speculating that the
radiation is the result of  bulk motion Comptonization of ambient
low-energy thermal photons by a cold conical wind with bulk Doppler
factor exceeding $10^7$ \citep{2002A&A...384..834A}.

In summary, independent of the details of the model for the EBL, the
HEGRA observations of H1426+428 between 700~GeV and 10~TeV are strongly
affected by absorption of the EBL between $1-15~\mu$m. The possible
strong absorption combined with the observable flux at TeV energies
implies an intrinsic spectrum which dominates the broad-band SED of
H1426+428.  This is not in agreement with the conceptual view of the
TeV-Blazars being the end-point of a phenomenological sequence where the
synchrotron component is comparable or dominates over the Compton component. The position
of the putative Compton peak is possibly at energies beyond 10~TeV.  As
has been shown here, the intrinsic source spectrum strongly depends upon
the shape of the SED of the EBL.  The different models of the EBL may be
excluded by requiring the intrinsic spectrum to be explainable in the
framework of conventional source models.

\begin{acknowledgements} 

The support of the German ministry for Research
and technology BMBF and of the Spanish Research Council CICYT is
gratefully acknowledged.  We thank the Instituto de Astrof\'{\i}sica de
Canarias for the use of the site and for supplying excellent working
conditions at La Palma. We gratefully acknowledge the technical support
staff of the Heidelberg, Kiel, Munich, and Yerevan Institutes. 

\end{acknowledgements}

\end{document}